\documentclass[preprint,12pt]{elsarticle}

\usepackage{graphics}
\usepackage{epsfig}
\usepackage{fancyhdr}
\usepackage{amssymb}
\usepackage{amsmath}

\begin{document}

\title{Multiplicity in pp and AA collisions: the same power law from energy-momentum constraints in string production}

\author[lis,san]{Irais Bautista}
\ead{irais@fpaxp1.usc.es}
\author[lis]{Jorge Dias de Deus}
\ead{jorge.dias.de.deus@ist.utl.pt}
\author[lis,cern]{Jos\'e Guilherme Milhano}
\ead{guilherme.milhano@ist.utl.pt}
\author[san]{Carlos Pajares}
\ead{pajares@fpaxp1.usc.es}

\address[lis]{CENTRA, Instituto Superior T\'ecnico, Universidade T\'ecnica de Lisboa, \\ Av. Rovisco Pais, P-1049-001 Lisboa, Portugal}
\address[cern]{Physics Department, Theory Unit, CERN, CH-1211 Gen\`eve 23, Switzerland}
\address[san]{IGFAE and Departamento de F\'isica de Part\'iculas, Univ. of Santiago de Compostela, 15782, Santiago de Compostela, Spain}

\begin{abstract}
We show that the dependence of the charged particle multiplicity on the centre-of-mass energy of the collision is, in the String Percolation Model, driven by the same power law behavior in both proton-proton and nucleus-nucleus collisions. The observed different growths are a result of energy-momentum constraints that limit the number of formed strings at low energy. Based on the very good description of the existing data, we provide predictions for future high energy LHC runs.

\end{abstract}

\maketitle

\section{Introduction}
\label{sec:intro}

Measurements of particle multiplicities constrain the early time properties of colliding systems. In the nucleus-nucleus case, these measurements are an essential ingredient for the estimation of the initial energy and entropy densities and thus the initial conditions from which the system will eventually thermalize and the quark gluon plasma be formed.

Data collected at RHIC and the LHC for proton-proton (pp) and nucleus-nucleus (AA) collisions establish unambiguously that nucleus-nucleus collisions are not a simple incoherent superposition of collisions of the participating nucleons,  $\left. {dn_{AA}}/{d\eta}\right|_{\eta=0}>N_{part} \cdot\left.{dn_{pp}}/{d\eta}\right|_{\eta=0}$, and thus that multiple scattering plays an important role.
Further, the possible scaling with the number of nucleon-nucleon also does not hold $\left. {dn_{AA}}/{d\eta}\right|_{\eta=0}\ll N_{coll} \cdot\left.{dn_{pp}}/{d\eta}\right|_{\eta=0}$, indicating that coherence effects among the relevant degrees of freedom at the nucleon level are at play during the collision process. 

The collision centre-of-mass energy dependence of the charged particle multiplicity in both pp and AA collisions is well reproduced by a power law as suggested by models, e.g. the Colour Glass Condensate or the String Percolation Model, where coherence effects play an important role. The logarithmic growth consistent with pre-LHC data is ruled out. 
Notably, this dependence is stronger \cite{Aamodt:2010pb} in the AA ($\left.{dn_{AA}}/{d\eta}\right|_{\eta=0}\sim {s}^{0.15}$) case than for pp ($\left.{dn_{pp}}/{d\eta}\right|_{\eta=0}\sim {s}^{0.11}$). 
Several possible explanations for this difference in energy dependence, naively at odds with theoretical expectations, have been put forward recently \cite{Levin:2011hr,Baier:2011ap,Lappi:2011gu,Rezaeian:2011ss,Tribedy:2011aa,Capella:2011vi}.

The String Percolation Model (SPM) \cite{Amelin:1993cs}, the Dual Parton model \cite{Capella:1992yb} including parton saturation effects, describes consistently properties of bulk multiparticle production \cite{DiasdeDeus:2007wb,Dias de Deus:2007tf,Brogueira:2007ub,Brogueira:2009ut,Bautista:2009my,Bautista:2010yf,Bautista:2011su}. This framework is closely related to descriptions based on the Colour Glass Condensate and Glasma flux tubes \cite{Lappi:2006fp} for which the same issues have been discussed extensively \cite{Kharzeev:2004if,Albacete:2005ef,ALbacete:2010ad}.

In this short note, we detail how the  SPM can provide for a joint description of  the energy dependence of multiplicity in pp and AA collisions once energy-momentum conservation constraints are taken into account. Further,  we show the resulting dependences on the nuclear species $A$ and number of participating nucleon $N_{\rm part}$ to be fully compatible with available data. 
Finally, we give predictions for future higher energy LHC runs and discuss generic expectations at high energy.

\section{Mid-rapidity multiplicity in the SPM}

The Glauber model and its generalizations (see \cite{Miller:2007ri} for a review) relate nucleus-nucleus collisions to collisions of their constituent nucleons. In the single scattering limit the average number of participating nucleons (per nucleon) $N_{A}=N_{\rm part}/2$ behave incoherently and the mid-rapidity multiplicities in nucleus-nucleus and proton-proton collisions are simply related by:
\begin{equation}
\label{eq:dndyAApplow}
	{\frac{dn^{AA}}{dy}}\bigg|_{y=0} \sim N_{A} \cdot \frac{dn^{pp}}{dy}\bigg|_{y=0} \, .
\end{equation}
This result, which corresponds to the wounded nucleon model \cite{Bialas:1976ed}, is expected to dominate at sufficiently low centre-of-mass energies.
At higher energies, multiple scattering becomes important, and 
\begin{equation}
\label{eq:dndyAApphigh}
	\frac{dn^{AA}}{dy}\bigg|_{y=0} \sim \big(N_{A}^{4/3} -N_A\big) \cdot\frac{dn^{pp}}{dy}\bigg|_{y=0} \, .
\end{equation}
Here, $N_{A}^{4/3}$ is the total number of nucleon-nucleon collisions and single scattering has been subtracted \cite{Armesto:2000xh,DiasdeDeus:2000cg}.

Energy-momentum conservation constrains the combinatorial factors of the Glauber model at low energy. In the framework of SPM, these constraints translate into the sharing of energy-momentum of $N_A$ valence strings amongst  $N_A^{4/3}$ (mostly sea) strings. A possible solution to this problem, the reduction of the height of the plateau for sea strings, was pursued in \cite{Armesto:2000xh,DiasdeDeus:2000cg}. Here, we proceed differently and account for energy-momentum conservation by reducing the effective number of sea strings via
\begin{equation}
	N_{A}^{4/3}\rightarrow N_{A}^{1+\alpha(\sqrt{s})}\, ,
\end{equation}
with
\begin{equation}
\label{eq:alpha}
	\alpha (\sqrt{s})=\frac{1}{3}\Bigg(1-\frac{1}{1+\ln(\sqrt{s/s_{0}}+1)}\Bigg)\, .
\end{equation}

We thus can write 
\begin{equation}
\label{eq:dndyAAppcorr}
	\frac{dn^{AA}}{dy} \bigg|_{y=0} \sim N_A \big(N_{A}^{\alpha (\sqrt{s})} - 1\big) \cdot\frac{dn^{pp}}{dy}\bigg|_{y=0} \, ,
\end{equation}
such that the wounded nucleon model eq.~(\ref{eq:dndyAApplow}) is recovered for $\sqrt{s}\ll \sqrt{s_{0}}$, and Glauber result eq.~(\ref{eq:dndyAApphigh}) follows for $\sqrt{s} \gg \sqrt{s_{0}}$, $\alpha(\sqrt{s}) \rightarrow \frac{1}{3}$.

In the SPM one considers Schwinger strings, which can fuse and percolate \cite{Armesto:1996kt,Braun:2000hd,Braun:2001us,Braun:2001su}, as the fundamental degrees of freedom. Multiparticle production is described in terms of these colour strings which are formed in the collision and  stretch between partons in the parting nuclei and are thus longitudinally extended in rapidity. %%%%%

The multiplicity of produced particles $dn/dy$ is proportional to the average number of such strings (twice the number of elementary collisions) $N^s$.  Thus, for a generic $N_A N_A$ collision, be it $pp$ or $AA$, one has
\begin{equation}
\label{eq:dndy1}
	\frac{dn}{dy} \bigg|_{y=0} \sim N^s_{N_A}\, .
\end{equation}

In the impact parameter plane, the colour content of the strings is confined within a small transverse area $S_1 = \pi r^2_0$, with $r_0\sim 0.2\div0.3$ fm. The strings decay via $q\bar{q}$ and $qq-\bar{q}\bar{q}$ pair production and subsequently hadronize to the observed hadrons. In the impact parameter plane, the strings appear as disks and with increasing energy-density these disks overlap, fuse and percolate leading to a reduction of the overall  charge \cite{Braun:1999hv}.

A cluster of $n$ strings behaves as a single string with energy-momentum
corresponding to the sum of the individual strings and with a higher colour field corresponding to the vectorial sum in colour space of the colour fields of the strings. In this way,
the mean multiplicity $\langle\mu_{n}\rangle$ and the mean transverse momentum squared $\langle p^{2}_{T}\rangle$ 
of the particles produced by a cluster in the limit of random distribution of strings
are given by $\langle\mu_{n}\rangle=N^{s}F(\eta^t)\langle\mu_{1}\rangle$ and $\langle p^{2}_{T}\rangle=\langle p^{2}_{T,1}\rangle/F(\eta^t)$ where $\langle\mu_{1}\rangle$ and $\langle p^{2}_{T,1}\rangle$ are the corresponding quantities for a single string. For a random distribution of disks, the colour reduction factor $F(\eta^t)$ is 
\begin{equation}
\label{eq:feta}
	F(\eta^{t})=\sqrt{\frac{1-e^{-\eta^{t}}}{\eta^{t}}}\, ,
\end{equation}
where $\eta^t$ is the impact parameter transverse density of strings (disks)
\begin{equation}
\label{eq:eta}
	\eta^{t} \equiv \frac{S_1}{S_{N_A}}\, {N}^{s} = \frac{\pi r^2_0}{S_{N_A}}{N}^{s}\, ,
\end{equation}
with $S_{N_A}$ the area of the impact parameter projected overlap of the interaction. For a $N_A N_A$ collision, eq. (\ref{eq:dndy1}) can be recast as 
\begin{equation}
\label{eq:dndy2}
	\frac{dn}{dy} \bigg|_{y=0} \sim F(\eta^t)\, N^s_{N_A}\, , \qquad N^s_{N_A} = N^s_{p} N_A^{1+\alpha (\sqrt{s})}\, .
\end{equation}
The colour reduction factor $F(\eta^t)$ results in a slowdown of the increase of $dn/dy$ with energy and number of participating nucleons. 
Note that if nucleons were to act incoherently, as in the limiting case eq. (\ref{eq:dndyAApplow}), $S_{N_A}$ would be given by the area of a single nucleon $S_p$, while once coherence is accounted for, eq. (\ref{eq:dndyAApphigh}), $S_{N_A}$ is the overall area of interaction. In general, $S_{N_A}$ depends on the impact parameter $b$ of the collision ($0\le b\le 2 R_A$) with $N_A\rightarrow 0, S_{N_A}\rightarrow 0$ as $b\rightarrow R_A$ and $N_A\rightarrow A, S_{N_A}\rightarrow S_A\equiv \pi R_A^2 \equiv \pi R^2_p A^{2/3}$ as $b\rightarrow 0$. These constraints are satisfied by 
\begin{equation}
	\frac{S_{N_{A}}}{S_{A}}=\bigg(\frac{N_{A}}{A}\bigg)^{\beta}\, ,
\end{equation}
with $\beta>0$. It follows that 
\begin{equation}
	S_{N_{A}}=\pi R^{2}_{p}\,A^{2/3}\bigg(\frac{N_{A}}{A}\bigg)^{\beta}\, ,
\end{equation}
and 
\begin{equation}
	\eta^{t}_{N_{A}}=\eta^{t}_{p}N_{A}^{\alpha(\sqrt{s})}A^{1/3}\bigg(\frac{A}{N_{A}}\bigg)^{\beta-1}\, .
\end{equation}
Here, motivated by the scaling limit of the number of vertices in a loop-erased random walk \cite{R.Keyon}, we set $\beta=5/3$ such that
\begin{equation}
	\eta^{t}_{N_{A}}=\eta^{t}_{p}N_{A}^{\alpha(\sqrt{s})}\bigg(\frac{A}{N_{A}^{2/3}}\bigg)\, .
\end{equation}

From eqs.~(\ref{eq:dndyAApplow}), (\ref{eq:dndyAApphigh}), (\ref{eq:dndyAAppcorr}), (\ref{eq:dndy2}) we can write
\begin{equation}
\label{eq:dndyfinal}
	\frac{1}{N_{A}}\, \frac{dn}{dy}\bigg|_{y=0} 
	= \frac{dn^{pp}}{dy}\bigg|_{y=0} \bigg(1+\frac{F(\eta^{t}_{N_{A}})}{F(\eta^{t}_{p})}(N^{\alpha(\sqrt{s})}_{A}-1)\bigg)\, ,
\end{equation}
The dependence of the multiplicity on the centre-of-mass collision energy $\sqrt{s}$ is fully specified once the average number of strings in a pp collision $N_p^s$ is known. At low energy $N_p^s$, is approximately equal to 2, growing with energy as $(\sqrt{s}/m_p)^{2\lambda}$ such that 
\begin{equation}
\label{eq:nsp}
	N_p^s = 2 + 4\bigg(\frac{r_0}{R_p}\bigg)^2 \bigg(\frac{\sqrt{s}}{m_p}\bigg)^{2\lambda}\, ,
\end{equation}
with $R_p$ the proton radius.

\section{Data description}

The energy and number of participants dependences implied by eq. (\ref{eq:dndyfinal}) can be tested against the available experimental data. Before doing so, two remarks on the multiplicity given by eq. (\ref{eq:dndyfinal}) are in order: (i) it is for all particles, while experimental data accounts only for charged particles; (ii) it is for mid-rapidity $y\sim 0$, while data is collected at mid pseudo-rapidity $\eta\sim 0$ and although these coincide, the multiplicity value should be rescaled by Jacobean of the $y\rightarrow \eta$ transformation. Neglecting the dependence of $p_T$, on which the Jacobean depends, on centre-of-mass energy and the (small) difference  between the Jacobean in pp and AA cases, both rescaling factors above can be absorbed into an overall normalization constant $\kappa$. We thus write the charged particle multiplicity at mid-rapidity in the pp case as
\begin{equation}
\label{eq:dndetapp}
	 \frac{dn^{pp}_{\rm ch}}{d\eta}\bigg|_{\eta=0} = \kappa\, F(\eta^{t}_{p})\, N_p^s\, ,
\end{equation}
such that, for the general $N_A N_A$ case, one has 
\begin{equation}
\label{eq:dndetaAA}
	\frac{1}{N_A} \frac{dn^{N_AN_A}_{\rm ch}}{d\eta}\bigg|_{\eta=0} =  \kappa\, F(\eta^{t}_{p})\, N_p^s \bigg(1+\frac{F(\eta^{t}_{N_{A}})}{F(\eta^{t}_{p})}(N^{\alpha(\sqrt{s})}_{A}-1)\bigg)\, .
\end{equation}

Eq.~(\ref{eq:dndetaAA}) depends on three parameters: (i) the normalization $\kappa$; (ii) the threshold scale $\sqrt{s_0}$ in $\alpha({\sqrt{s}})$ in eq.~(\ref{eq:alpha}); (iii) the power $\lambda$ in eq.~(\ref{eq:nsp}).
We performed a global fit to pp data \cite{UA1ppm,UA51,STAR,CDF,Aamodt:2010ft,Khachatryan:2010xs,Khachatryan:2010us} in the range $ 53\le\sqrt{s}\le 7000$ GeV, and to AA (AuAu, CuCu and PbPb) at different centralities \cite{Alver:2010ck,Aamodt:2010cz} for $19.6\le\sqrt{s}\le 2760$ GeV. The fitted data sample consists of 116 points (19 for pp and 97 for AA).

The best fit yields the parameter values: $\kappa = 0.63\pm0.01$, $\lambda=0.201\pm0.003$, and $\sqrt{s_0}=245\pm29$ GeV. One notices that while $\kappa$ and $\lambda$ are tightly constrained, the threshold scale $\sqrt{s_0}$ is determined with a sizable associated error of $\sim$ 10\%. This results from the mild dependence of eq.~(\ref{eq:dndetaAA}), through eq.~(\ref{eq:alpha}), on $\sqrt{s_0}$ compounded with the sizable errors in the existing measurements (see Fig.~\ref{fig2} below).

Fig.~\ref{fig1} shows a comparison of the evolution of the mid-rapidity multiplicity with energy given by eq.~(\ref{eq:dndetaAA}) with data for pp and AA collisions.

\begin{figure}[h]
\centering
\includegraphics[angle=0,width=\linewidth]{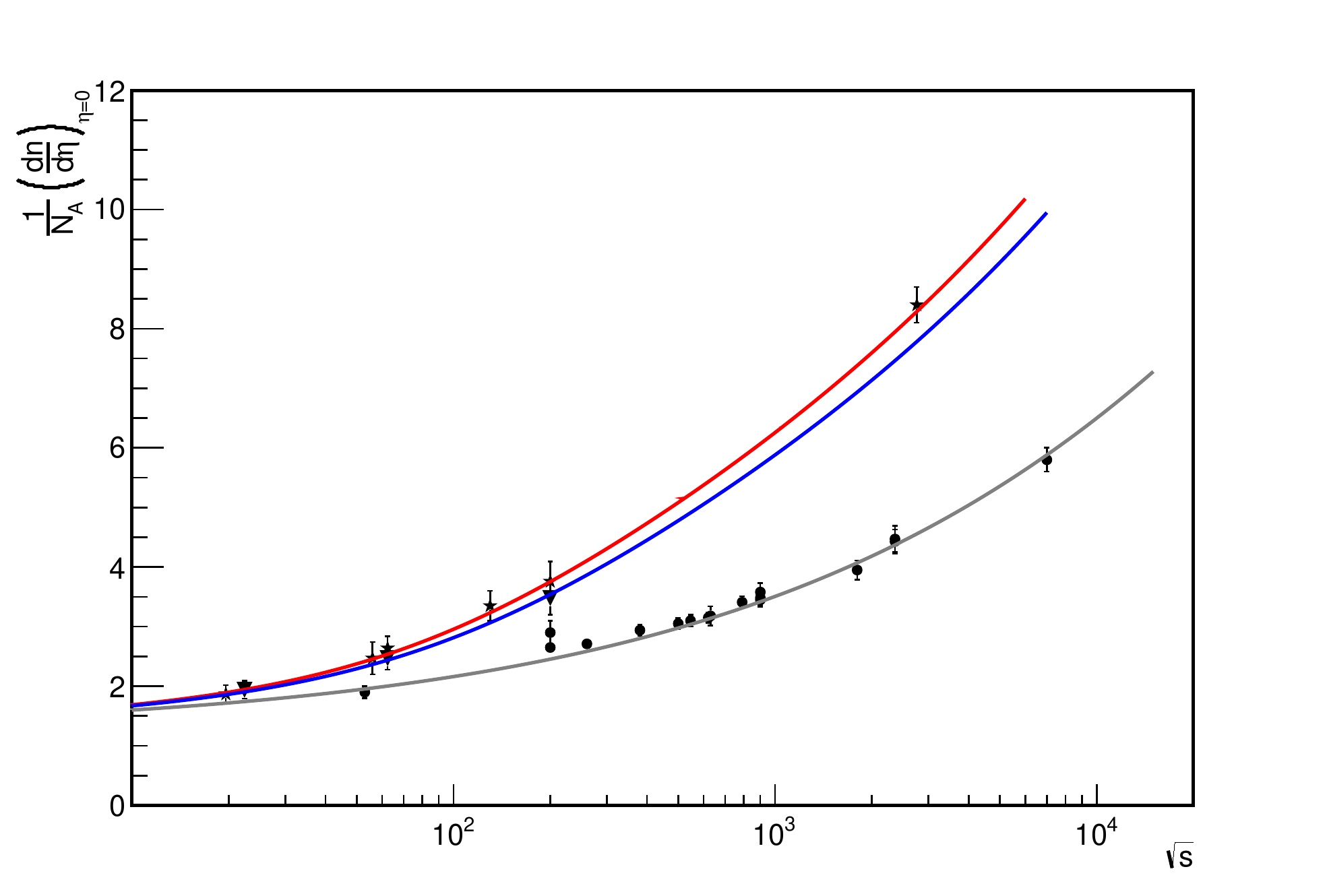}
\caption{\label{fig1}
Multiplicity dependence on $\sqrt{s}$. pp data from \cite{UA1ppm,UA51,STAR,CDF,Aamodt:2010ft,Khachatryan:2010xs,Khachatryan:2010us} (circles), CuCu (triangles) and AuAu (stars) from \cite{Alver:2010ck}, PbPb (star) from \cite{Aamodt:2010cz}. Curves obtained from eq.~(\ref{eq:dndetaAA}): ($N_A=1$,  $A=1$) for pp (grey line); ($N_A=50$,  $A=63$) for CuCu (blue line); and ($N_A=175$,  $A=200$) for AuAu/PbPb (red  line). Color online. }
\end{figure}

For reference we quote, in Table \ref{tab1}, the predicted values relevant for future LHC runs.

\begin{table}[h]
\begin{center}
\scriptsize
\begin{tabular}{c|c|c|c||c|c|c|}
%\hline
\cline{2-7}
  & \multicolumn{3}{||c||}{pp} &  \multicolumn{3}{|c|}{PbPb} \\
% \hline
 \cline{1-7}
\multicolumn{1}{|c|}{$\sqrt{s}$ (TeV)} &\multicolumn{1}{||c|}{8}		&  	10   	&	14	 &	3.2	&	3.9	&	5.5 \\
\hline
 \multicolumn{1}{|c|}{$1/N_A\cdot\left.{dn^{N_AN_A}_{\rm ch}}/{d\eta}\right|_{\eta=0}$} &	\multicolumn{1}{||c|}{6.10$\pm$0.14}	&  	
 {6.50$\pm$0.16}   	&	{7.14$\pm$0.18}	 &	{8.69$\pm$0.14}	&	{9.22$\pm$0.15}	&	{10.1$\pm$0.2} \\
\hline
\end{tabular}
\normalsize
\end{center}
\caption{Predicted multiplicities for pp and PbPb at future LHC energies.}
\label{tab1}
\end{table}%

The dependence of the multiplicity on the number of participating nucleons implied by eq.~(\ref{eq:dndetaAA}) is compared with data in Fig.~\ref{fig2}. 

\begin{figure}[h]
\centering
\includegraphics[angle=0,width=\linewidth]{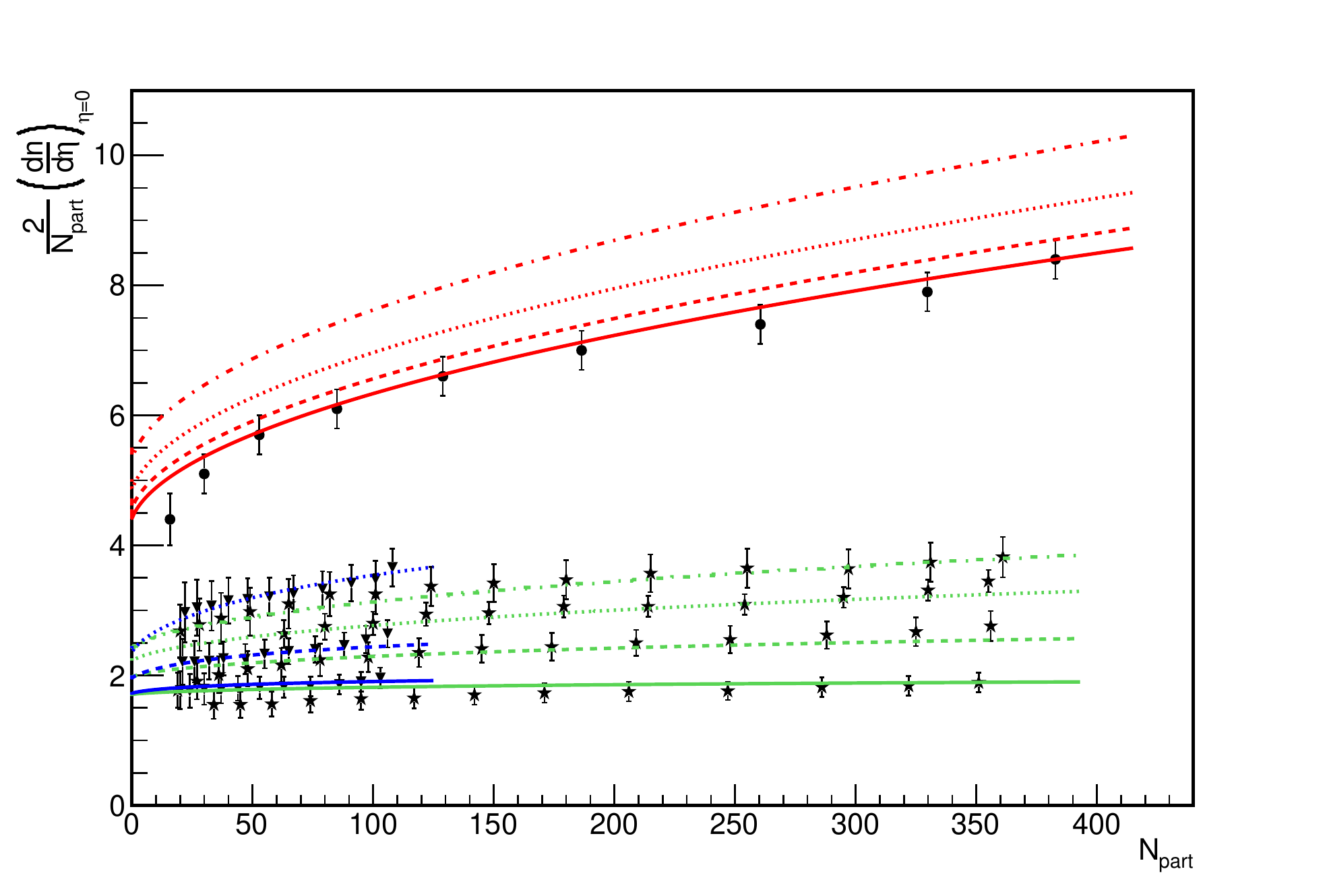}
\caption{\label{fig2}
Multiplicity dependence on centrality (the number of participating nucleons $N_{\rm part}=2 N_A$ where $N_A$ is the number of participants per nucleus). CuCu (triangles) and AuAu (stars) data from \cite{Alver:2010ck}, and PbPb (circles) from  \cite{Aamodt:2010cz}. Curves obtained from eq.~(\ref{eq:dndetaAA}): ($\sqrt{s}=22.4, 62.4, 200$ GeV) for CuCu (blue); ($\sqrt{s}=19.6, 62.4, 130, 200$ GeV) for AuAu (green); and ($\sqrt{s}=2.76, 3.2, 3.9, 5.5$ TeV) for PbPb (red). Color online. }
\end{figure}

\section{Conclusions}

We have shown that, in the SPM, the power law dependence of the multiplicity on the collision energy is the same in pp and AA collisions. The slower growth in the AA case is due to finite energy-momentum constraints which tamper string creation, and thus the multiplicity, at low energy. 
In the high energy limit, $F(\eta^{t})\rightarrow (\eta^t)^{-1/2}$ and $\alpha(\sqrt{s})\rightarrow 1/3$,  eq.~(\ref{eq:dndetaAA}) can be written 
\begin{equation}
\label{eq:dndyhe}
	 \frac{1}{N_{A}}\,\frac{dn^{N_AN_A}_{\rm ch}}{d\eta}\bigg|_{\eta=0} =   \frac{dn^{pp}_{\rm ch}}{d\eta}\bigg|_{\eta=0} \Bigg(1+\bigg(\frac{N_{A}}{A} \bigg)^{1/2} \bigg(1 - \frac{1}{N_A^{1/3}}\bigg)\Bigg)\, .
\end{equation}
With the obtained fit parameters, this asymptotic result becomes a good approximation (within 5\% of that given by eq.~(\ref{eq:dndetaAA})) for 
$\sqrt{s}\gtrsim 500$ GeV. This indicates that finite energy corrections persist to fairly high energies.

In this high energy limit, the shape of  $\frac{dn^{N_AN_A}}{d\eta}$ as a function of the number of participants is energy independent, that is to say that the $N_{\rm part}$ and $\sqrt{s}$ dependences factorize. In the SPM, energy-momentum conservation results in violations of this factorization and they are the origin of the observed discrepancy in multiplicity growth with energy in pp and AA.

The arguments put forward in this short note can be readily adapted to the case of asymmetric (proton-nucleus) collisions.  Also in this case, energy-momentum constraints are expected to play an important role.

\section*{Acknowledgments}
We thank N. Armesto for useful discussions.
IB is supported by  the grant SFRH/BD/51370/2011 from Funda\c c\~ao para a Ci\^encia e a Tecnologia (Portugal).
JDD and JGM acknowledge the support of Funda\c c\~ao para a Ci\^encia e a Tecnologia (Portugal) under project CERN/FP/116379/2010. IB and CP were partly supported by the project FPA2008-01177  and FPA2011-22776 of MICINN, the Spanish Consolider Ingenio 2010 program CPAN and Conselleria de Educacion Xunta de Galicia.

%J. D. D. thanks the support of the FCT/Portugal project PPCDT/FIS/575682004. 

\end{document}